# APPENDIX F

**Scaling Behaviour in the Number of Criminal Acts Committed by Individuals**

**Will Cook, Paul Ormerod* and Ellie Cooper**

**Volterra Consulting**

**October 2003**

* Corresponding author  pormerod@volterra.co.uk

We are very grateful to Professor David Farrington of the University of Cambridge and Rebecca Stallings of the University of Pittsburgh for supplying us with and helping us understand their data sets.  All interpretations are our own.

*Abstract*

*We examine the distribution of the extent of criminal activity by individuals in two widely cited data bases. The Pittsburgh Youth Study measures self-reported criminal acts over intervals of six months or a year in three groups of boys in the public school system in Pittsburgh, PA. The Cambridge Study in Delinquent Development records criminal convictions amongst a group of working class youths in the UK over a 14 year period.*

*The range of the data is very substantially different between these two measures of criminal activity, one of which is convictions and the other self-reported acts. However, there are many similarities between the characteristics of the data sets.*

*A power law relationship between the frequency and rank of the number of criminal acts describes the data well in both cases, and fits the data better than an exponential relationship. Power law distributions of macroscopic observables are ubiquitous in both the natural and social sciences. They are indicative of correlated, cooperative phenomena between groups of interacting agents at the microscopic level.*

*However, there is evidence of a bimodal distribution, again in each case. Excluding the frequency with which zero crimes are committed or reported reduces the absolute size of the estimated exponent in the power law relationship. The exponent is virtually identical in both cases. A better fit is obtained for the tail of the distribution.*

*In other words, there appears to be a subtle deviation from straightforward power law behaviour. The description of the data when the number of boys committing or reporting zero crimes are excluded is different from that when they are included. The crucial step in the criminal progress of an individual appears to be committing the first act. Once this is done, the number of criminal acts committed by an individual can take place on all scales.*

# *1.     Introduction*

An increasing number of social and economic phenomena have been shown to exhibit power law distributions in aspects of their behaviour. Power law distributions of macroscopic observables are ubiquitous in both the natural and social sciences. They are indicative of correlated, cooperative phenomena between groups of interacting agents at the microscopic level. Examples include the world wide web [1, for example], sexual contacts [2], the size of firms [3] and the extinction rates of firms [4].

However, in other instances, whilst power laws may provide a good description, there may be subtle deviations from such scaling behaviour. Ref [5], for example, raises the possibility of exponential truncation of power law behaviour in terms of the world wide web. The duration of economic recessions under capitalism is well described by a power law, but there are deviations from this behaviour when one considers the frequency of occurrence of large recessions [6]. Ref [7] goes so far as to suggest that many phenomena which approximate power law behaviour may possibly be better described by variants of an exponential relationship.

In this paper, we examine the distribution of the extent of criminal activity by individuals in two widely cited data bases.

The first, the Cambridge Study in Delinquent Development, is a prospective longitudinal survey of 411 males in a working class area of North London in the UK. Data collection began in 1961-62. The second, the Pittsburgh Youth Study began in 1986 with a random sample of boys in the first, fourth, and seventh grades of the Pittsburgh, PA, public school system. The sample contains approximately 500 boys at each grade level, for a total of 1,517 boys.

Most crime is committed by young men, and both the Cambridge and the Pittsburgh studies monitor behaviour over time in groups of youths. The Cambridge data which we analyse relates to the number of convictions for each boy over a period spanning the mid 1960s and 1970s. The Pittsburgh data we analyse describes self-reported acts of delinquency over short time intervals beginning in the late 1980s.

The data we have selected from each study therefore differ both in their time coverage and in the fact that the Cambridge study describes convictions for offences and the Pittsburgh one describes self-reported acts of delinquency. This enables us to examine different aspects of criminal behaviour at different times.

## 2. *The data*

The Cambridge Study in Delinquent Development is a prospective longitudinal survey of 411 males in a working class area of North London in the UK. Data collection began in 1961-62 when most of the boys were aged 8-9. The Cambridge Study is a virtually unique data base of criminal activity.

It is a prospective longitudinal survey over a period of over 20 years, and the focus of interest is crime and delinquency. Many variables were measured before the youths were officially convicted, to avoid the problem of retrospective bias. The study involved frequent personal contacts with a group of boys and their parents, so records were supplemented by interview, tests and questionnaire data. At least up to age 21, there was a very low attrition rate.

From the outset, a fairly representative sample of urban working class youths was followed up, rather than extreme groups of (predicted or identified) delinquents and non-delinquents, so that all degrees of delinquency were present. The officially delinquent minority became gradually differentiated from their non-delinquent peers, avoiding the problem of selection of control groups.

The sample was limited to males from a working class urban area because of the prior expectation of a high prevalence of convictions (about a quarter) among them. The sample size was set at about 400 because this was considered large enough for statistical comparisons between convicted and unconvicted boys, whilst also being small enough to interview each boy and his family and build up intensive case histories.

On the basis of their fathers' occupations, 93.7 per cent could be described as working class (categories III, IV, or V on the Registrar General's scale of occupational prestige), in comparison with the national figure of 78.3 per cent at that time. This was, therefore, overwhelmingly a white, urban, working class sample of British origin.

Data is available for the total number of times each boy was convicted of a criminal offence between the ages of 10 and 24. More precisely, the data is available for 395 out of the total original sample of 411. Convictions were only counted if they were for offences normally recorded in the Criminal Record Office, which are more or less synonymous with "serious" or "criminal" offences. For example, no convictions for traffic offences were included, nor convictions for offences regarded as minor (e.g. public drunkenness or common assault). The most common offences included were thefts, burglaries, and unauthorized takings of motor vehicles.

The Pittsburgh Youth Study began in 1987 with a random sample of boys in the first, fourth, and seventh grades of the Pittsburgh, PA, public school system. Information from the initial screening was used to select the top 30 percent of boys with the most disruptive behaviour. This group of boys, together with a random sample of the remaining 70 percent who showed less disruptive behaviour, became the sample for the study. The sample contains approximately 500 boys at each grade level, for a total of 1,517 boys.

Each student and a primary caregiver were interviewed at 6-month intervals for the first 5 years of the study; teacher ratings of the student were also obtained. The middle sample (fourth grade) was discontinued after seven assessments. The youngest sample (first grade) and oldest sample (seventh grade) are currently being interviewed at annual intervals, with totals of 16 and 14 assessments, respectively. The study has been highly successful in retaining participants, with a retention rate of at least 85 percent for each assessment. Interviews are not continuing at this time; the last assessment of the oldest sample was in 2000 and the youngest in 2001. Data is available for each of the time intervals described above for the number of self-reported acts of delinquency committed by each boy.

There are issues in the construction of the Pittsburgh data relating to the handling of missing observations. These can arise because the team were unable to interview a particular individual at the relevant time, or because the interviewee either refused to answer or could not remember how many things he had done. Essentially, a participant in the study is registered as having committed acts of delinquency in the relevant period if his interview is missing some answers but less than 30 per cent of the total. In this case, the mean of the non-missing answers is substituted for the missing answer. Readers with a natural science background may recoil in horror at such procedures, but problems such as these are entirely routine in data collection in the social sciences.

## 3. The results

### 3.1 The Cambridge study

In the Cambridge study, in total, 130 out of the 395 boys in the sample were convicted on at least one occasion, and the maximum number of convictions was 14, recorded for 2 of the boys.

Figure 1 shows the log-log plot of the frequency with which each number of convictions was observed and the rank of the number. The rank is used so that zero convictions can be used in the analysis. Zero convictions has rank of 1, one conviction rank of 2, and so on.

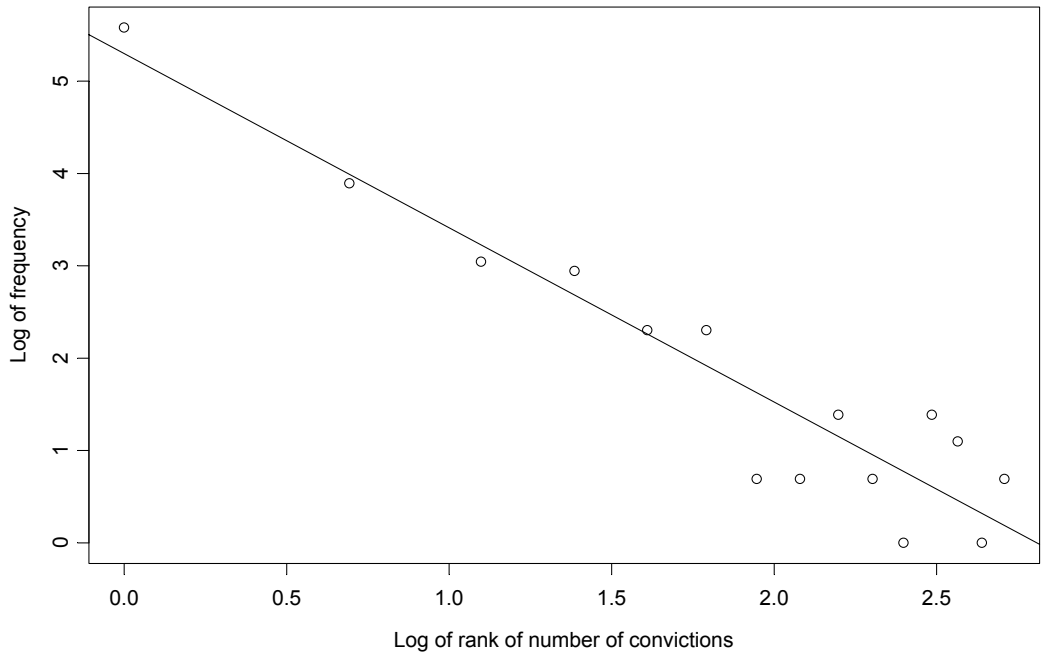

**Figure 1** *Cambridge Study in Delinquent Development. Frequency of total number of convictions of each of 395 boys between ages 10-24 and rank of the number (zero has rank of 1, one conviction has rank of 2, and so on).*

This suggests that the data follow a power law distribution. We formalized the analysis by use of least-squares regression on the relationship

$$F = \alpha.N^{\beta} \qquad (1)$$

where F is frequency and N is rank.

Direct non-linear least squares estimation of (1)[1] gave a better fit overall than a log-log transformation. The estimated value of β is –2.28 with a standard error of 0.071. The standard error of the equation is 3.46.

Table 1 shows the actual values of frequency and those fitted by the least-squares regression.

---

[1] using the non-linear least squares algorithm in the package S-Plus

**Table 1      Actual frequency and frequency fitted by least-squares regression**

| Number of convictions | Actual | Fitted |
|---|---|---|
| 0 | 265 | 264 |
| 1 | 49 | 54 |
| 2 | 21 | 22 |
| 3 | 19 | 11 |
| 4 | 10 | 7 |
| 5 | 10 | 4 |
| 6 | 2 | 3 |
| 7 | 2 | 2 |
| 8 | 4 | 1 |
| 9 | 2 | 1 |
| 10 | 1 | 1 |
| 11 | 4 | 1 |
| 12 | 3 | 1 |
| 13 | 1 | 1 |
| 14 | 2 | 1 |

*Note:* fitted values rounded to nearest whole number

The power law relationship is clearly a good description of the data, although there does tend to be slightly more of the higher numbers of convictions than the regression predicts. However, equation (1) fits the data much better than an exponential relationship between frequency and rank, with the respective equations standard errors being 11.6 and 26.8 per cent respectively of the mean of the data.

A better fit to the more extreme values can be obtained by focusing purely on the frequency of the number of boys with actual convictions. In other words, excluding those with no convictions. The estimated value of β is lower in absolute value, at

−1.18 with a standard error of 0.076. The standard error of the equation is 2.47. Table 2 shows the actual and fitted values of this equation.

**Table 2**     **Actual frequency and frequency fitted by least-squares regression, excluding zero convictions from the sample**

| Number of convictions | Actual | Fitted |
| --- | --- | --- |
| 1 | 49 | 50 |
| 2 | 21 | 22 |
| 3 | 19 | 14 |
| 4 | 10 | 10 |
| 5 | 10 | 7 |
| 6 | 2 | 6 |
| 7 | 2 | 5 |
| 8 | 4 | 4 |
| 9 | 2 | 4 |
| 10 | 1 | 3 |
| 11 | 4 | 3 |
| 12 | 3 | 3 |
| 13 | 1 | 2 |
| 14 | 2 | 2 |

*Note:* fitted values rounded to nearest whole number

Again, the power law relationship provides a better description of the data than the exponential, with the latter being unable to capture the length of the tail in the actual data.

*3.2    The Pittsburgh study*

The Pittsburgh study relates not to actual criminal convictions but to self-reported acts of delinquency. The scale of the data is therefore quite different. The Cambridge data ranges from zero to 14 convictions, all obtained over a period of 14 years. In contrast, the average number of self-reported acts of delinquency during the course of at most one year across the Pittsburgh data is around 30. Excluding those who report zero acts of delinquency the average is just over 90, with a very small number of boys claiming over 1,000 such acts.

There is, of course, a massive distinction between being convicted and simply committing a criminal act, and some of the Cambridge convictions may well have been for considerably more than one act of delinquency. A common phrase in court reporting in the UK is that someone found guilty 'asked for N other offences to be taken into consideration'.

Interestingly, however, the proportion of boys in each time period who are crime free is virtually the same in both data sets, being some two-thirds.

The data set is divided into those boys in the first grade at the beginning of the study, those in the fourth grade and those in the seventh grade. It is further divided into the different phases, with interviews being conducted at periods of either six months or one year. In total, there are 23 separate sets of data which can be analysed.

The average number of boys for which data is available across the 23 separate data sets is 421. The majority of boys report no acts of delinquency in the relevant period. Of those who do report such acts, 87 per cent report a number of between 1 and 20. Given that the highest number reported is in each data set more than 500 and in most of them more than 1000, there is obviously a very long tail to the typical data set. In other words, non-zero frequencies are recorded for most frequencies between 1 and 20, but there are subsequently long strings of frequencies with value zero, punctuated by the occasional individual who reports a very large number of delinquent acts.

However, a comparison of the least squares fits of $F = \alpha.N^{\beta}$ both including and excluding the frequencies taking the value of zero showed that the estimated parameters of the equation are virtually identical in both cases across each of the data sets. This is not surprising, since least-squares estimation will by definition attempt to fit the bulk of the data, which occurs at relatively low levels of reported acts of delinquency. We report below the results with the zero frequency observations excluded.

The most important point to note about these results is their similarity to those of the Cambridge study. This is despite the facts that they record events over quite different time intervals, and that one study records convictions whereas the other records self-reported acts of delinquency.

A power law relationship provides a better description of the data sets than does an exponential relationship, though the difference is not as strong as it is with the Cambridge data. The average equation standard error across the 23 Pittsburgh data sets is 2.787 for the power law of equation (1) and 3.412 for an exponential fit.

The average estimate of β in the Pittsburgh data sets is –3.52 compared to –2.28 in the Cambridge data. When the boys who report no crime in the relevant period are excluded from the estimation, in both cases the absolute value of the estimated value of β falls. With the Cambridge data it is –1.18 and the average with the Pittsburgh data is –1.00

Table 3 summarises the estimated values of β.

**Table 3**  Estimated values of β in equation 1: $F = \alpha . N^{\beta}$

|  | Including zero convictions/crimes | Excluding zero convictions/crimes |
|---|---|---|
| Cambridge data | -2.28 | -1.18 |
| Average of Pittsburgh data sets | -3.52 | -1.00 |
| Pittsburgh maximum (absolute) | -5.44 | -1.23 |
| Pittsburgh minimum (absolute) | -2.17 | -0.85 |

Again, as with the Cambridge data, the exclusion of observations with zero reported crime improves the fit of the equation in the tail. Tables 4a and 4b respectively set out a typical example of the fitted values from equation (1) with a Pittsburgh data set, including and excluding the zero value.

**Table 4a**   Actual and fitted values of equation 1:  $F = \alpha.N^{\beta}$
Including zero reported crimes: typical example of a Pittsburgh data set

| Number of reported crimes[1] | Actual Frequency | Fitted Frequency |
|---|---|---|
| 0 | 288 | 286 |
| 1 | 52 | 64 |
| 2 | 30 | 27 |
| 3 | 17 | 14 |
| 4 | 17 | 9 |
| 5 | 13 | 6 |
| 6 | 9 | 4 |
| 7 | 11 | 3 |
| 8 | 4 | 2 |
| 9 | 4 | 2 |
| 10 | 5 | 2 |
| 11 | 3 | 1 |
| 12 | 6 | 1 |
| 13 | 0 | 1 |
| 14 | 5 | 1 |
| 15 | 0 | 1 |
| 16 | 3 | 1 |
| 17 | 3 | 1 |
| 18 | 2 | 0 |
| 19 | 6 | 0 |
| 20 | 1 | 0 |

*Note: Numbers 0 to 20 only are reported in the table. This range accounts for 95 per cent of the total number of boys in this sample*

**Table 4b**     Actual and fitted values of equation 1:  $F = \alpha.N^{\beta}$

Excluding zero reported crimes: typical example of a Pittsburgh data set

| Number of reported crimes | Actual Frequency | Fitted Frequency |
|---|---|---|
| 1 | 52 | 55 |
| 2 | 30 | 26 |
| 3 | 17 | 17 |
| 4 | 17 | 12 |
| 5 | 13 | 10 |
| 6 | 9 | 8 |
| 7 | 11 | 7 |
| 8 | 4 | 6 |
| 9 | 4 | 5 |
| 10 | 5 | 5 |
| 11 | 3 | 4 |
| 12 | 6 | 4 |
| 13 | 0 | 4 |
| 14 | 5 | 3 |
| 15 | 0 | 3 |
| 16 | 3 | 3 |
| 17 | 3 | 3 |
| 18 | 2 | 2 |
| 19 | 6 | 2 |
| 20 | 1 | 2 |

## *4.     Conclusion*

An increasing number of social and economic phenomena have been shown to exhibit power law distributions in aspects of their behaviour. Power law distributions of macroscopic observables are indicative of correlated, cooperative phenomena between groups of interacting agents at the microscopic level. However, in other instances, whilst power laws may provide a good description, there may be subtle deviations from such scaling behaviour. In particular, there may be a bimodal distribution of the observed macroscopic data, with a power law fitting part of the data better than it does the whole.

In this paper, we examine the distribution of the extent of criminal activity by individuals in two widely cited data bases. The first, the Cambridge Study in Delinquent Development measures is a prospective longitudinal survey of young males in a working class area of North London in the UK. We use data on the total number of convictions per individual from this study. The second, the Pittsburgh Youth Study is a sample of boys in the first, fourth, and seventh grades of the Pittsburgh, PA, public school system. We use data from this study on the self-reported number of acts of delinquency over periods of six months and a year.

We therefore examine different aspects of criminal activity - convictions and self-reported acts of delinquency - over different time-scales.

The results are remarkably similar in the two separate studies. A power law relating the frequency to the rank of the number of convictions/reported criminal acts fits the overall samples well, and better than the alternative hypothesis of an exponential relationship between the two.

However, the exclusion of the frequency with which zero crimes are committed/reported improves the fit with both data sets. In other words, when the sample is restricted to those individuals who have actually committed a crime in the relevant time period, a power law describes the frequency distribution even better than when the whole sample is used.

When zero crimes are excluded, the exponent of the power law relationship is virtually the same value when both the Cambridge and the Pittsburgh data sets are analysed.

The results suggest that the distribution of convictions and criminal offences is bimodal. In particular, there is an important distinction between those who do not commit a criminal act in a given time period, and those who do. The crucial step in the criminal progress of an individual appears to be committing the first act. Once this is done, the number of criminal acts committed by an individual can take place on all scales.